# Molecular Adsorption of $H_2O$ on $TiO_2$ and $TiO_2$:Y Surfaces

D. D. Nematov [1, 2*], K.T. Kholmurodov [3, 4], M. A. Husenzoda [2], A. Lyubchyk [5, 6], A. S. Burhonzoda [1, 2]

[1] *S.U. Umarov Physical-Technical Institute of the National Academy of Science of Tajikistan, Dushanbe, Tajikistan*

[2] *Osimi Tajik Technical University, 724000, Dushanbe, Tajikistan*

[3] *Joint Institute for Nuclear Research, Dubna, Moscow Region, 141980, Russia*

[4] *Dubna State University, Dubna, Moscow Region, 141980, Russia*

[5] *Nanotechcenter LLC, Krzhizhanovsky str. 3, 03680 Kyiv, Ukraine*

[6] *Lusófona University, IDEGI, Campo Grande, 376 1749-024, Lisboa, Portugal*

**Abstract**

In this work, using theoretical calculations within the framework of the density functional theory, taking into account the dispersive VDW interaction, the processes of adsorption and interaction of a water molecule with a $TiO_2$ surface in various configurations are investigated. At the atomic / molecular level, the interactions of a water molecule with a $TiO_2$ surface have been studied for various orientations. The results of calculations within the framework of DFT + VDW show that the adsorption energies of single water molecules in different initial positions on the substrate surface vary from -0.72 to -0.84 eV, and the most stable adsorbate structure is the $TiO_2 + H_2O$ system upon adsorption of a molecule of water, parallel to the Y axis, because during the adsorption of $H_2O$ parallel to the Y axis, some favorable effects are observed in the band structure of titanium dioxide. On the one hand, the band gap decreases to 2.59 eV, and on the other hand, a new energy state appears in the band gap with an energy contribution of 0.17 eV, when water is physisorbed and interacts with a titanium atom at a distance of 2.12 Å and occupies a perpendicular position relative to the surface.

## 1. Introduction

With the ongoing threat of the energy crisis and global warming caused by the increase in the use of fossil energy, the search for sustainable and environmentally friendly sources of energy is one of the most urgent challenges of human civilization [1, 2], since it is known that the continued use of fossil fuels in the world threatens our energy supply and creates a huge burden on the environment. Research on the use of sustainable green energy represents one of the ways to mitigate the growing threat of global environmental problems and the energy crisis, which is very intense and active around the world. Solar panels and wind turbines have become familiar to us. However, new advances in nanotechnology and materials science make it possible to collect energy from other sources and will allow and implement the creative idea of Nikola Tesla about "Getting an electric current from the air". Recently, scientists and engineers have been working on the creation of innovative devices for converting humidity into electricity, which will expand the range of known renewable energy sources due to a new source of atmospheric humidity (galvanic converters that convert air humidity into electricity). That is, such devices are capable of collecting electricity from atmospheric humidity and supplying electrical current, similar to how solar panels capture sunlight and generate electricity.

It is known that moisture is ubiquitous on Earth (71% of the Earth's surface is covered with water), containing a huge reservoir of low-potential energy in the form of gaseous water molecules and water droplets. It was found that a number of functional nanomaterials such as $TiO_2$, $CaSiO_3$, $ZrO_2$, $SnO_2$, $Al_2O_3$, as well as biofibers and carbon materials can generate electricity directly when interacting with moisture [3-5].

This indicates the possibility of generating electrical energy from atmospheric moisture, and allows the creation of self-powered devices. While this technology is still evolving, there are already some strategies for improving the energy conversion efficiency and power output in these devices. Various materials for energy conversion, including carbon nanoparticles, graphene, metal oxides, biofibers and polymers have been investigated for their efficiency and among them $TiO_2$, $ZnO$, and $ZrO_2$ have shown the generated voltage varies from 2.5 to -1.6 V depending on the humidity of the environment [6].

In our previous work, the issues of interaction and adsorption of the $H_2O$ molecule on the surface of $ZrO_2$ + 3 mol% $Y_2O_3$ were studied and it was found that water has a significant effect on the electronic-band structures and it was shown that the surface of $ZrO_2$ nanoparticles can be considered as a reaction zone for electrochemical processes [7]. An important observation in this work was that the introduction of 3% mol of $Y_2O_3$ doping into the $H_2O$ / $ZrO_2$ system led to a shift in the DOS amplitude from 4 eV to 0. This shows that water molecules are adsorbed from the atmosphere on the surface of $ZrO_2$-based nanoparticles leads to the realization the process of localization of an electron from the crystal lattice of nanoparticles and its transfer outside the particles. In another work Droshkevich A.S. et al. [8] reported on the direct conversion of the energy of adsorption of water into electricity on the surface of zirconia nanoparticles when doped with 3 mol% $Y_2O_3$.

Titanium dioxide nanoparticles have many properties in common with $ZrO_2$ nanoparticles. The similarity of these materials is mainly observed in the band structure, electron recombination, chemical properties, etc. [9-10]. While there is such a strong similarity in the band structure and electronic properties of these materials, it can be assumed that their transport and chemisorption properties are close to each other. $TiO_2$ with a band gap in the 3-3.3 eV limit [11] is an important material with a wide range of applications, for example, in photocatalysts, heterogeneous catalysts, adsorption hydroelectric converters (moisture converters), electronic devices, production of sensitized solar cells, water splitting, bone implants, sensors, etc. [12-14]. On the other hand, the effect of air humidity on the hexagonal phase of titanium dioxide particles remains unclear, although $TiO_2$ nanowires with a negative surface charge were the first nanostructures of metal oxides used to create electricity from moisture [15]. Since nanowires are negatively charged, cations in water (mainly protons) can migrate together with water molecules through nanochannels between nanowires during diffusion of moisture into the nanowire network [15]. It has been shown that a moisture-controlled electric generator based on the diffusion flow of water in nets of $TiO_2$ nanowires can provide an output power density of up to 4 μW/cm² when exposed to a very humid environment [16]. In this case, these devices are completely autonomous.

Molecular and dissociative adsorption of chemical compounds can significantly affect the electronic and surface properties of $TiO_2$. $H_2O$ molecules are among the most important substances causing these effects [17]. Experimental work shows that $H_2O$ can significantly improve the CO oxidation turnover frequency [18, 19]. Hence, understanding the adsorption of $H_2O$ and its interaction on the $TiO_2$ surface is necessary in order to know the internal mechanism. It is known that, under normal conditions, $TiO_2$ has oxygen vacancies [20], and these defects are considered important coordinates for adsorption and stimulation of many surface reactions. $H_2O$ - molecules can easily eliminate defects in oxygen vacancies through dissociation. It is reported that the dissociation of water, on the other hand, leads to the formation of OH groups on the $TiO_2$ surface [21]. Such OH groups can significantly affect the electronic properties of $TiO_2$, which can lead to further adsorption and diffusion of $H_2O$ and, as a result, a wet electronic state and interband electron transfer are formed due to the interaction of surface atoms with water molecules [21]. The aim of this study is to theoretically study the adsorption of a water molecule on a $TiO_2$ surface using density functional theory (DFT) taking into account dispersion interactions.

## 2. Calculation Methodology

Interaction processes and adsorption of molecules and surfaces can be studied using experimental methods such as scanning tunneling microscopy (STM), atomic force microscopy (AFM), as well as theoretical methods such as density functional theory (DFT) and molecular dynamics modeling (MD). We calculated the total energy for the interaction of the $H_2O$ molecule with the $TiO_2$ surface in the framework of the DFT, using the Wien2k package [22]. The Kohn – Sham equations [23] were solved to describe the states of the electron nucleus. The electron exchange energies and correlations were calculated within the Perdew-Burke-Ernzerhof (PBE) generalized gradient approximation (GGA) [24]. For the accuracy of calculations of the total energy and adsorption energy, dispersive VDW interactions were also taken into account by the semi-empirical Grimm method (DFT-D2) in the Wien2k package, since it is known that interatomic interactions are ignored at distances exceeding a certain radius. All calculations are spin polarized. The Kohn-Sham single-electron states were expanded in accordance with the basis sets of plane waves with a kinetic energy of 400 eV. The energy of adsorption (Eads) was calculated and corrected as the difference between the total energy of the system and the sum of the energy, which is calculated separately for each structure that makes up the adsorption structure according to the recommendation [25].

$$E_{ads}^{corrected} = -[totE_{surf+H_2O} - (totE_{surf} + totE_{H_2O})] \qquad (1)$$

Thus, in order to obtain the adsorption energy of a water molecule, we had to calculate separately the total energy for the following three optimized systems:

    a. Supercells $TiO_2$ and $TiO_2$:Y (adsorbent)

    b. $H_2O$ molecules (adsorbate)

    c. $TiO_2$: Y + $H_2O$ and $TiO_2$ + $H_2O$ systems (adsorption system)

An example algorithm for calculating the adsorption energy of a water molecule on the surface of titanium dioxide is shown below (Figure 1).

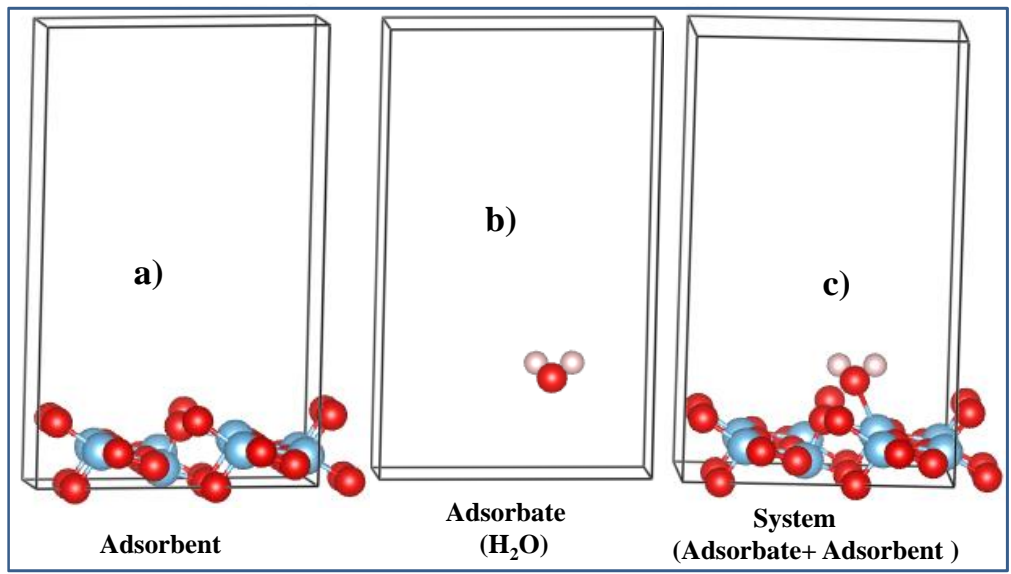

**Figure 1. Structural images of the titanium dioxide cell, water molecules and water on the titanium dioxide surface. Titanium atoms are blue, oxygen atoms are red, and hydrogen atoms are white**

The total energy of these systems within the DFT, taking into account the energy of long-range van der Waals (vdW) attraction arising from the electron correlations of the adsorbate ($H_2O$) - substrate was calculated as follows:

$$\left(\frac{\hbar^2}{2m}\nabla^2 + E_{ion}(r) + E_h(r) + E_{xc}^{vdW+DF}(r) + E_{vdW}(r)\right)\psi_i(r) = E\,\psi_i(r) \qquad (2)$$

where energy exchange and correlation of electrons adsorbate and subtrate - ($E_{xc}^{vdW+DF}(r)$) determined according to the methodology for determining the VdW interaction within the framework of the DFT [26, 27]. Here $E_{ion}(r)$ - the potential describing the ionic interaction), and $E_{vdW}(r)$ – potential of van der Waals (vdW) interactions.

Energy $E_{vdW}(r)$ defined as:

$$E_{vdW}(r) = \frac{C_{vdW}}{(r - r_{vdW})^3} f(2k_c(r - r_{vdW})) \qquad (3)$$

where $r_{vdW}$ and $C_{vdW}$ respectively are the radius of action of the VdV forces and the position of the van der Waals plane (vdW), depending on the dielectric properties of the substrate and the polarizability of the adsorbate [28, 29].

To carry out the calculations, the $TiO_2$ unit cells were first optimized. Then, based on the optimized unit cells, supercells doped with Y were created to reduce the time of quantum-chemical calculations of the adsorption process, since it is known that the adsorbate breaks and reduces the symmetry in the system. All calculations were carried out taking into account the spin-orbit and spin-polarized effects of electron interaction.

## 3. Results and Discussion

Section According to our calculations, the band gap of a clean $TiO_2$ surface is 3.2 eV, which is in good agreement with the existing experimental data (3-3.23 eV) [30-32] and indicates the correctness of our quantum-chemical calculations and the correctness of the problem statement within the DFT framework. While DFT methods are not good at estimating bandgap values, the calculations taking into account the VdW interaction in this study give better agreement with experiment than previous theoretical results [33]. On the other hand, Y causes minor defect states in the valence bands within the band gap, which prevents hybridization of Y (3d) and unoccupied oxygen 2p orbitals, and the band gap decreases to 3.12 eV. The results of calculating the electronic properties for pure titanium dioxide in the form of DOS graphs, which are presented in this work (Figure 2 (a)) have much in common (with the exclusion of their energy forbidden zone) with nanocrystals of zirconium dioxide in undoped form (Figure 2 (b)) from the point of view of the structure of band structures and energy levels at the edges of the valence band and the origins of the conduction bands, although, according to previous works [34, 7], the band gap of $ZrO_2$ is greater than that of titanium dioxide. In both systems, for the formation of valence energy states, electrons of oxygen atoms mainly play a role, and metal atoms (Zr and Ti) contribute to the formation of energy states at the beginning of the conduction band. Figure 2 (a, b) shows similar graphs of the total density of electronic states of these materials in the energy range -5 eV to 5 eV, which are important for understanding the contribution of each atom and specific electronic states to the formation of the valence band and the conduction band near Fermi levels.

The enhanced density of states for $TiO_2$ means an increase in vacancies in their outer electron orbitals, that is, there are many places available to occupy. It can be seen that, due to the wide band gap typical of the three pure $TiO_2$ structures - anatase, rutile, and brookite (3.0–3.3 eV), insufficient conversion of solar energy and low quantum efficiency are characteristic as a result of the high recombination rate of photogenerated electron-hole pairs. Overcoming these disadvantages and modifying the electronic band structure of $TiO_2$ by doping with ions of transit metals, surface adsorption by organic molecules, the formation of heterojunctions or with other semiconductors with a lower band gap, such as metal oxides and chalcogenides, can be carried out to improve the characteristics of $TiO_2$ [35, 36]. Under these conditions, in most cases, either intermediate energy states arise between the valence band (VB) and the conduction band (CB), or the process leads to a narrowing of the $TiO_2$ band gap. For example, surface adsorption of organic molecules occurs through two different mechanisms.

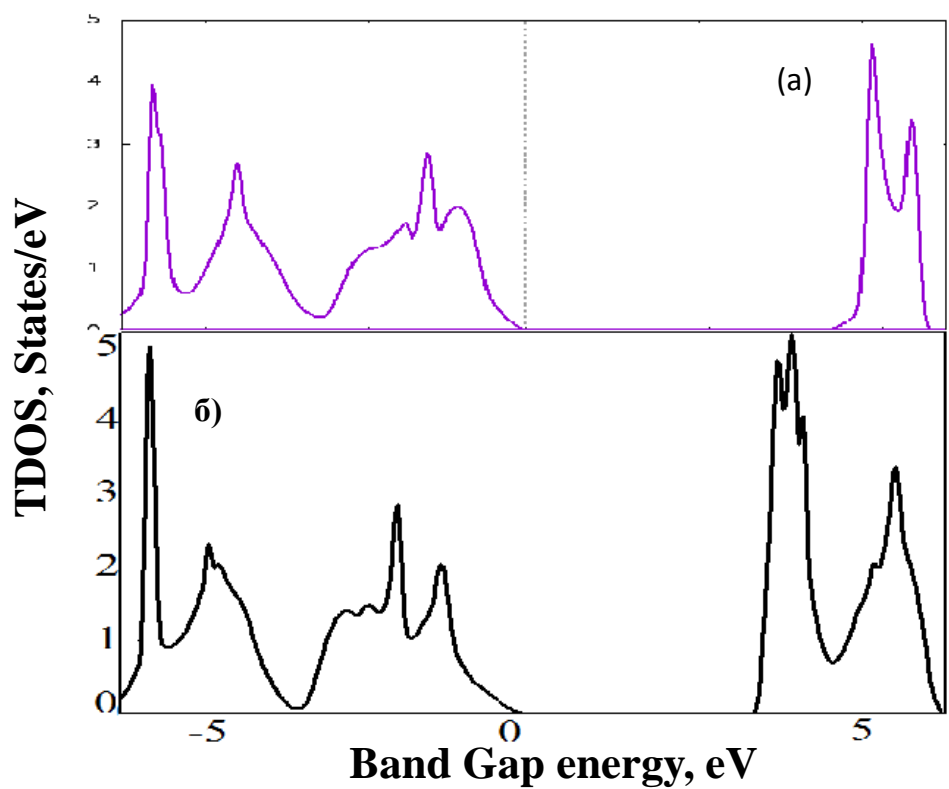

**Figure 2. Densities of electronic states: a) for $ZrO_2$ [34], b) for pure $TiO_2$. Fermi level approaches point 0 (b) and dotted lines in (a)**

In the first case, large molecules are adsorbed onto the $TiO_2$ surface, and electrons are injected from the excited molecule into the conduction band of the semiconductor. In the second, relatively small molecules adsorbed on the surface form a charge transfer complex, which is absorbed in the visible region at energies lower than that of chelating molecules or substrates [35, 36].

At the next stage, in the same way, we identified the adsorption of a water molecule on the surface of pure and Y-doped $TiO_2$ nanocrystals at the same arbitrary adsorption site (all other coordinates are also possible binding sites for any orientation of water for certain distances) of the $TiO_2$ surface depending on their orientation relative to the X and Y axes. The center of mass of the oxygen molecule was used as a reference point for the initial position of the $H_2O$ molecule. The total energy of the static system was calculated and the adsorption energies of a water molecule on the substrate surface were found and are shown in Table 1.

Table 1. Energy adsorption of $H_2O$ - molecules on the $TiO_2$ surface

| System / $H_2O$ Orientation | Link length, Å | | Energy adsorption, eV | Bandgap, eV |
|---|---|---|---|---|
| | O-Ti | O-Y | | |
| $TiO_2$+ $H_2O$ / Parallel to X | 2.34 | - | −0.78 | 2.85 |
| $TiO_2$+ $H_2O$ / Parallel to Y | 2.12 | - | −0.72 | 2.59 |
| $TiO_2$:Y+ $H_2O$ / Parallel to X | - | 2.48 | −0.84 | 2.75 |
| $TiO_2$:Y+ $H_2O$ / Parallel to Y | - | 2.23 | −0.77 | 3.07 |

Our calculations show that in all cases, water is molecularly adsorbed on the substrate surface, and in vertical adsorption, water molecules are bound through oxygen O atoms over pure titanium dioxide atoms at a distance of d = 2.12 Å. In this mode, the $H_2O$ molecule is adsorbed with an energy of –0.72 eV. The other three geometries are very close in terms of the adsorption energy, and, therefore, they can be considered competitive to this system. Almost equally stable, with an adsorption energy of -0.77 eV, is a configuration when a molecule is vertically adsorbed on the surface of Y-tri-doped $TiO_2$ at a distance of d = 2.23 Å. When $H_2O$ approaches parallel to the surface of pure titanium dioxide, the adsorption energy is -0.78 eV at the nearest molecule distance to d = 2.34 Å (Figure 3).

### 3.1. Water Adsorption and Zone Structure of the Substrate

Understanding water / solid interactions is of great importance in the various fundamental properties of $TiO_2$. For example, when $H_2O$ interacts, the catalytic, electronic and surface properties of $TiO_2$ may change due to the adsorption of a water molecule. To assess the effect of adsorbed water molecules on the properties of $TiO_2$ adsorption, we analyzed the total density of states for various forms of water adsorption on the substrate surface. It was found that, in all cases, $H_2O$ is molecularly adsorbed on the substrate surface, although according to the literature, the adsorption process in most cases can lead to dissociation of the molecule on the substrate surface; however, this process depends on the distance at which the interaction between the adsorbent and the adsorbate occurs. On the other hand, the adsorption of $H_2O$ on the surface of titanium dioxide doped with titanium can cause an electron transfer from Y to $TiO_2$ and, accordingly, an interband transition of electrons in $TiO_2$, due to a decrease in the energy gap and the emergence of new energy states within the forbidden band. As expected, for some systems the main contribution below the edge of the valence band is made by the oxygen 2p orbitals of the $TiO_2$ surface, while above the edge of the conduction band, mainly 3d orbitals of titanium dioxide are present. It is not difficult to imagine that during the adsorption of a water molecule on the surface, occupied orbitals appear in the valence band due to these molecules. Figures 3-6 show the optimized structured images and DOS for all systems under study.

It can be seen that the band gap of titanium dioxide decreases to 2.85 eV, when a water molecule is adsorbed on the surface parallel to the X axis. At the same time, a sharp increase in the density of electronic states of the material is also observed, leading to the growth of defective vacancies in the system (Figure 3). Estimation of the adsorption energies of $H_2O$ shows that it is likely that at further approaches, water dissociates, and in this case, one H is transferred to the surface oxygen, and OH remains at the top of the Ti atom.

However, when molecules are adsorbed parallel to the Y axis, some other beneficial effects are observed in the band structure of titanium dioxide. On the one hand, the band gap decreases to 2.59 eV, and on the other hand, new shallow and deeply located energy states with an energy contribution of 0.17 eV appear inside the energy shell (Figure 4) when water is physisorbed and interacts with a titanium atom at a distance of 2.12 Å and occupies perpendicular to the surface.

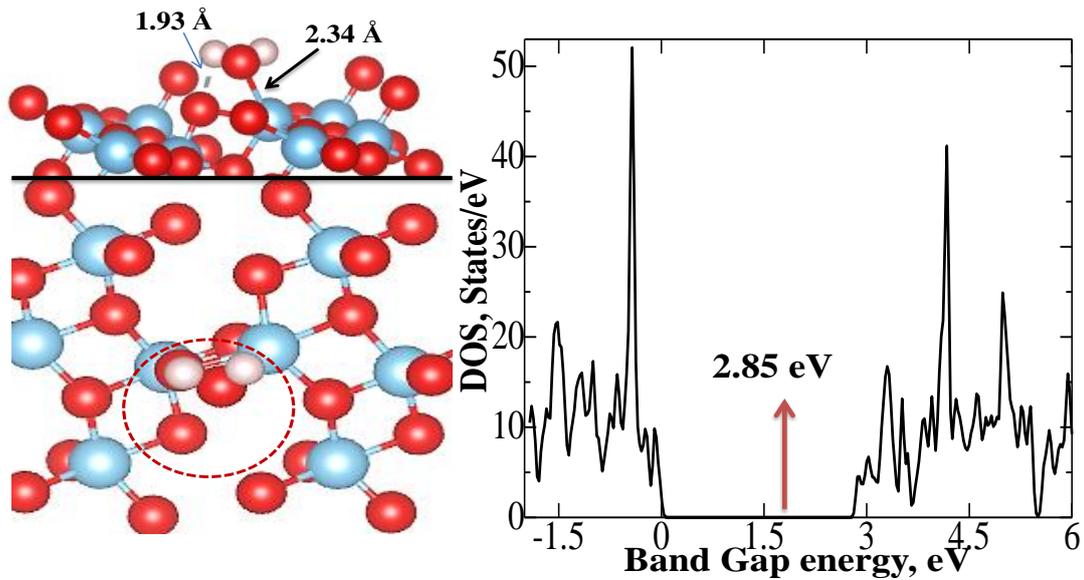

**Figure 3. Structural images of the top layer and DOS of the TiO$_2$ system during adsorption of H$_2$O - molecules parallel to the X axis**

The formation of such new energetic sprawling zones within the forbidden band plays an important role in the transport of charge through materials. Electrons of oxygen and titanium atoms in the substrate can naturally contribute to the narrowing of the band gap and the formation of new energy states (Figure 4); however, it is assumed that electrons of the atomic orbitals of the water molecule play a role in the formation of these states, during which surface reactions and charge transfer. Similar behavior of systems is also confirmed in previous works when studying the adsorption of water on the surface of Ni-doped ZrO$_2$ nanocrystals. In addition, it was found that water on the surface of titanium dioxide has a stronger adsorption energy (-0.72 eV), but it can be seen that despite the small close interaction, water is still molecularly adsorbed without signs of dissociation. An experimental study of the adsorption of H$_2$O molecules from the surface of hexagonal titanium dioxide has not yet been carried out, so we investigated the interaction of water on four corresponding surfaces without comparison with the results of experimental measurements.

At the next stage, the adsorption of water over the titanium dioxide titanium alloy surface was studied (Figure 5). For titanium dioxide doped with yttrium, the adsorption of a water molecule parallel to the X axis can lead to dissociation of H$_2$O. However, in the H$_2$O/Y + TiO$_2$ system, a new energy level is localized in the form of an interband peak, which possibly arises due to yttrium atoms. This facilitates the interaction between the yttrium atom and the oxygen orbitals of H$_2$O. In addition, the effect of yttrium on the electronic structure of the H$_2$O molecule, possibly, contributes to enhancing their chemisorption, since signs of a hydrogen bond are observed in the system.

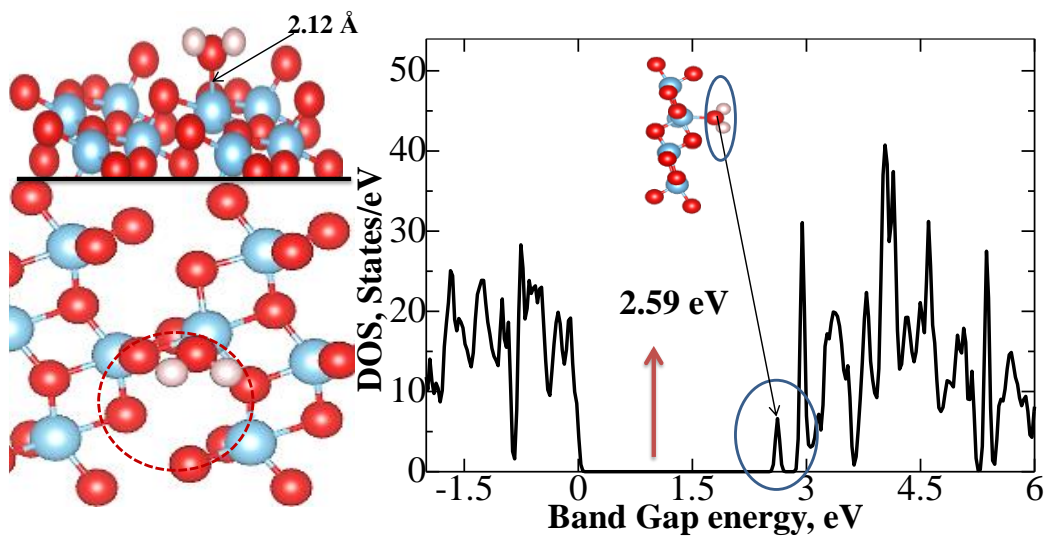

**Figure 4. Structural images of the upper layer and DOS of the TiO$_2$ system during adsorption of H$_2$O - molecules parallel to the Y axis**

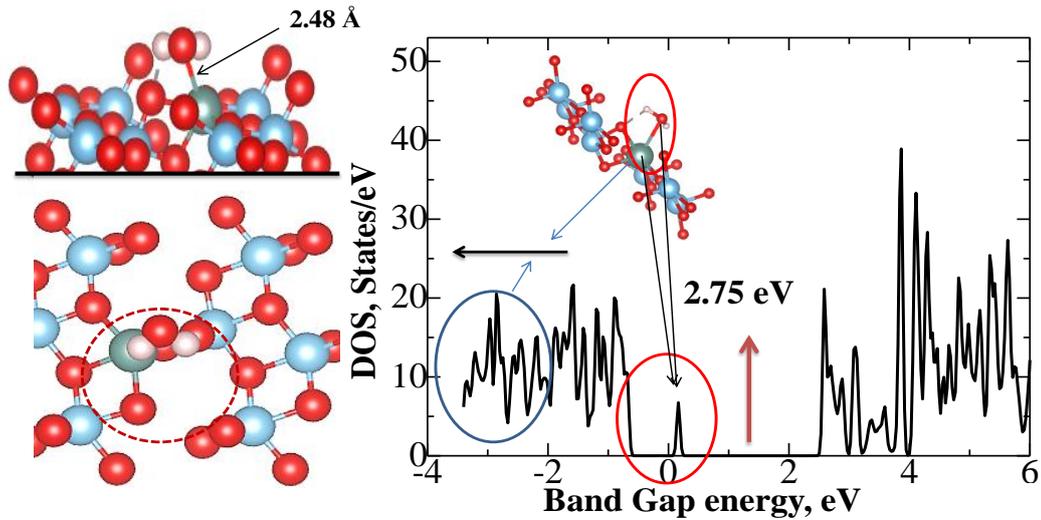

**Figure 5. Structural images of the top layer and DOS of the TiO$_2$: Y system upon adsorption of H$_2$O – molecules parallel to the X axis**

The perpendicular incidence of a water molecule on the surface of titanium dioxide doped with yttrium causes the band gap to decrease from 3.2 to 3.07 eV, and the molecules are adsorbed on the surface of titanium dioxide in molecular form interacting with at a distance of 2.23 Å (Figure 6). The energy of adsorption of a water molecule on the Y + TiO$_2$ surface with adsorption parallel to the X axis was exactly - 0.77 eV.

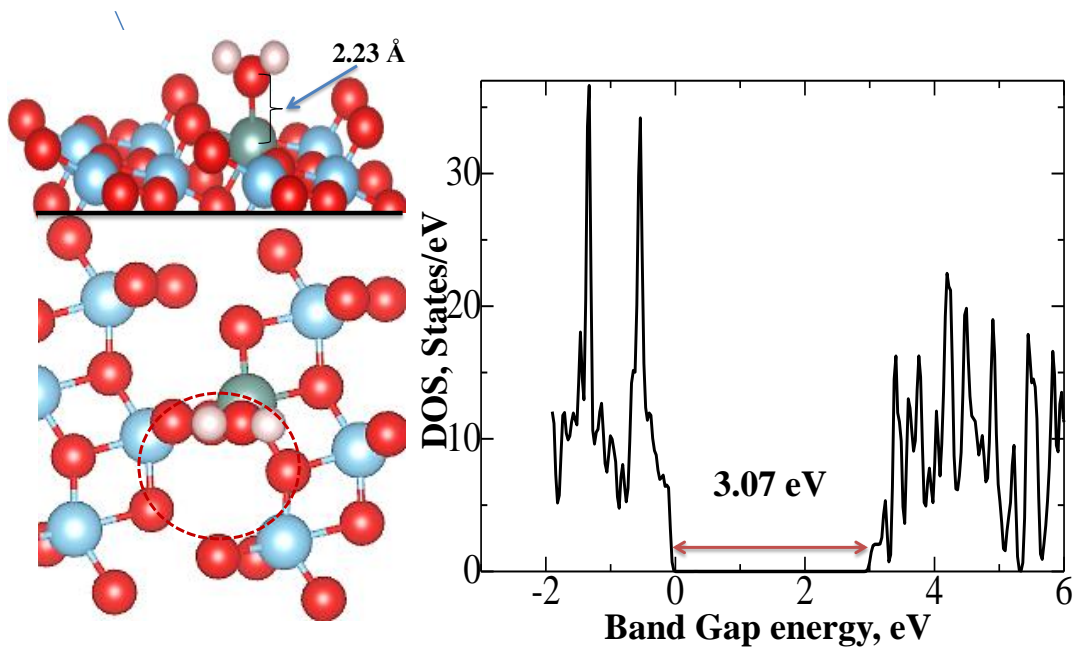

**Figure 6. Structural images of the top layer and DOS of the TiO$_2$:Y system upon adsorption of H$_2$O – molecules parallel to the Y axis**

The results of DFT calculations show that the adsorption energies of single water molecules at different initial positions on the substrate surface varied from –0.72 to –0.84 eV. It can be seen from the results presented that, depending on the orientation of the water and the distance of interaction of H$_2$O from the substrate surface, the system behaves differently. In some favorable cases, due to the influence of water, new energy states appear inside the forbidden band, which leads to the possibility of the transition of electrons to the conduction band. The difference in the electron density of adsorption systems in the presence of metal ions is well noticed, and it can be assumed that this leads to charge transfer and a change in the spatial position between the interacting water atoms and the surface of the substrate. Yttrium ions interact with the titanium dioxide surface through electrostatic interaction, subsequently affecting the adsorption of the water molecule and the surface hydration characteristics. In some cases, electrons can be transferred from the water molecule to the surface of titanium dioxide, as a result of which the surface may be recharged. Thus, we studied the adsorption of water molecules on the surface of titanium dioxide and $TiO_2$ doped with yttrium at the same arbitrary adsorption site depending on their orientation relative to the X and Y axes. For a detailed consideration of such processes, it is necessary to study the adsorption of water with all possible orientations as well as adsorption along various coordinates (surfaces of individual atoms, bridge regions, hollow places), since all other coordinates can also be associated with the surface for any orientation of water at certain distances. On the other hand, according to Lee et al. [36], the generation power and the output current strongly depend on the orientation of water vapor adsorption. The results obtained may contribute to the understanding of some features of these materials that are important for their practical application and may be of interest to researchers working in this field. The results can be used by other researchers to model the structure of substances that are supposed to be synthesized, as well as to determine such an important component as "*composition-structure-property*".

## 4. Conclusion

Applying exact exchange-correlation functionals in the framework of DFT, taking into account dispersion interactions, we were able to reveal that water in its molecular form is favorably adsorbed on the $TiO_2$ surface in all orientations. It can be concluded that the most stable adsorption with charge transfer can occur when $H_2O$ molecules lie flat on the surface of pure titanium dioxide or vertically on the surface of Y-doped $TiO_2$. The results of DFT calculations show that the adsorption energies of single water molecules at different initial positions on the substrate surface varied from –0.72 to –0.84 eV. At such favorable times, it can be assumed that the potential applied to the surface is large enough to induce chemisorption. It turned out that when a water molecule is adsorbed on the substrate surface, the band gap decreases and, in some cases, the Fermi levels noticeably shift. The perpendicular incidence of a water molecule on the surface of titanium dioxide doped with yttrium causes the band gap to decrease from 3.2 to 3.07 eV, and the molecules are adsorbed on the surface of titanium dioxide in molecular form, interacting with it at a distance of 2.23 Å. When a water molecule is adsorbed parallel to the Y axis inside the band gap of titanium dioxide and titanium dioxide doped with yttrium, new deep allowed energy levels with a weak energy contribution are formed, while when water falls on these surfaces parallel to the X axis, only a narrowing of the band gap is observed.

## 5. Declarations

### 5.1. Author Contributions

Conceptualization, D.N., M.H., A.L., and A.B.; methodology, D.N., M.H., K.K., and A.B.; software, D.N., and K.K.; formal analysis, D.N., M.H., A.L., K.K., and A.B.; writing—original draft preparation, D.N. and A.L.; supervision, M.H. and A.B. All authors have read and agreed to the published version of the manuscript.

### 5.2. Data Availability Statement

The data presented in this study are available on request from the corresponding author.

### 5.3. Funding

This publication is part of a project, SSHARE - Self-sufficient "humidity to electricity" innovative radiant adsorption system toward net zero energy buildings, that has received funding from the European Union's Horizon 2020 research and innovation program under grant agreement N°871284 for their sponsorship to realize this work.

### 5.4. Acknowledgements

### 5.5. Institutional Review Board Statement

Not Applicable.

### 5.6. Informed Consent Statement

Not Applicable.

### 5.7. Declaration of Competing Interest

The authors declare that there is no conflict of interests regarding the publication of this manuscript. In addition, the ethical issues, including plagiarism, informed consent, misconduct, data fabrication and/or falsification, double publication and/or submission, and redundancies have been completely observed by the authors.